\documentclass[aps,showpacs,groupedaddress,superscriptaddress,twocolumn]{revtex4}

\usepackage{amsfonts}
\usepackage{amsmath} 
\usepackage{amssymb}
\usepackage{graphicx}
\usepackage{latexsym}
\usepackage{makeidx}
\usepackage{color}
\usepackage{bm}
\usepackage{dcolumn}

\begin{document}
\preprint{}


\title{Orientation dependence of high-harmonic generation in monolayer transition metal dichalcogenides}
\author{Tomohiro Tamaya}
\altaffiliation{E-mail: t-tamaya@aist.go.jp}
\affiliation{Nanolectronics Research Institute (NeRI), National Institute of Advanced Industrial Science and Technology (AIST), Tsukuba, Ibaraki 305-8568, Japan}
\affiliation{CREST, Japan Science and Technology Agency, Kawaguchi, Saitama 332-0012, Japan}
\author{Satoru Konabe}
\affiliation{Research Institute for Science and Technology, Tokyo University of Science, Katsushika, Tokyo 125-8585, Japan}
\author{Shiro Kawabata}
\affiliation{Nanolectronics Research Institute (NeRI), National Institute of Advanced Industrial Science and Technology (AIST), Tsukuba, Ibaraki 305-8568, Japan}
\affiliation{CREST, Japan Science and Technology Agency, Kawaguchi, Saitama 332-0012, Japan}


\date{\today}


\begin{abstract}

We theoretically investigate the orientation dependence of high-harmonic generation (HHG) in monolayer transition metal dichalcogenides (TMDCs). We find that, unlike conventional solid-state and atomic layered materials such as graphene, both parallel and perpendicular emissions with respect to the incident electric field exist in TMDCs. Interestingly, the parallel (perpendicular) emissions principally contain only odd-(even-) order harmonics. Both harmonics show the same periodicity in the crystallographic orientations but opposite phases. These peculiar behaviors can be understood on the basis of the dipole moments in TMDCs, which reflect the symmetries of both atomic orbitals and lattice structures. Our findings are qualitatively consistent with recent experimental results and provide a possibility for high-harmonic spectroscopy of solid-state materials.

\end{abstract}



\maketitle


Atomically thin two-dimensional materials have been widely investigated in recent years because of their potential utilities for optoelectronic technologies. Transition metal dichalcogenides (TMDCs) are typical representatives \cite{Radisavijevic,Wang,Sanchez} having hexagonal lattices of $B$ ($B =$ Mo or W) and $A$ ($A =$ S or Se) atoms with inversion symmetry breaking (Fig. 1(a)). The inversion symmetry breaking gives rise to a bandgap energy at the $K^{\pm}$ points and provides a good platform for valley contrasting physics~\cite{Rycertz,Cao,Mak,Zeng,Mak3}. Valley contrasting physics yields a unique perspective on the optical properties of TMDCs~\cite{Spelendiani,Zhao}, including valley-dependent optical selection rules for interband excitations processes of Bloch electrons with a circularly polarized electric field \cite{Xiao, Jones, Xu, Sallen}. Thus, TMDCs may have useful applications in optoelectronics and may provide a means for investigating fundamental aspects of optics.


\begin{figure}[b] 
\includegraphics[width=8.5cm]{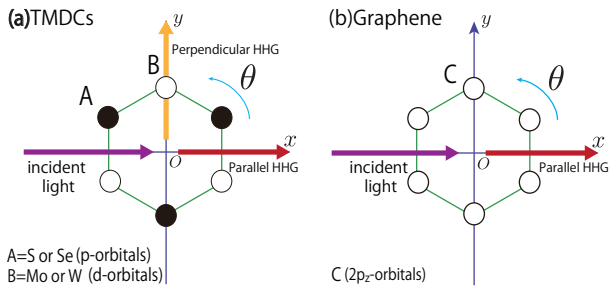}
\caption{Schematic diagrams of the crystallographic orientation dependence of HHG in (a) TMDCs and (b) graphene.}
\end{figure}

A fundamental topic in optics is high-harmonic generation (HHG) \cite{Boyd,Shen,Yariv}. Recent experiments in solid-state materials have promoted this phenomenon to the non-perturbative regime and revealed novel properties \cite{Ghimire,Schubert,Luu,Vampa,Hohenleutner,Langer}. HHG in TMDCs has also been intensely investigated because TMDCs are expected to have atypical light-matter interactions compared with ordinary semiconductors \cite{Kumar,Malard,Zeng2,Li,Liu,Wang3}. Some experiments have shown that HHG emissions in TMDCs are quite sensitive to the crystallographic orientations with respect to the incident electric field~\cite{Kumar, Malard, Zeng2,Li,Liu,Wang3}, which has also been observed in some special materials such as MgO and GaSe \cite{You,Langer2}. This characteristic of TMDCs has been explained only in terms of the symmetries in lattice structures and does not take into account the atomic orbitals in solid-state materials~\cite{Kumar,Malard,Li,Zeng2}. This consideration implies that there may be a possibility of developing high-harmonic spectroscopy of solid-state materials and this possibility could be confirmed by comparing materials with the same lattice structures and different atomic orbitals, such as TMDCs and graphene (Fig. 1).

Here, we theoretically investigate the orientation dependence of HHG in monolayer TMDCs and graphene by taking into account the symmetries of both crystal structures and atomic orbitals in solid-state materials. We show the orientation dependence of HHG in TMDCs and graphene to be quite different. This difference can be attributed to the wavenumber dependences of the dipole moments, which are fundamentally determined from both symmetries of atomic orbitals and lattice structures in solid-state materials.


To investigate the orientation dependence of HHG, we start from the tight-binding model, where two-dimensional hexagonal lattices $BA_2$ are constructed from $A$ ($A=$S or Se) and $B$ ($B=$Mo or W) atoms \cite{Mallic,Tatsumi}. Here, we assume the wavefunctions of $A$ and $B$ to be of the form $\phi_A(\bm{x})=p_x(\bm{x})+ i p_y (\bm{x})$ ($p$ orbitals for the chalcogen atoms) and $ \phi_B(\bm{x})=d_{xy} (\bm{x})+i \tau^z d_{x^2-y^2} (\bm{x}) $ ($d$ orbitals for metal atoms), respectively, where $\tau^z=\pm1$ is the variable describing the state at the $K^\pm$ points. This assumption is only valid near $K^\pm$ points. We also assume that the difference between the onsite energies for the atoms is $m=\varepsilon_A-\varepsilon_B$.

\begin{figure}[bp] 
\includegraphics[width=8.5cm]{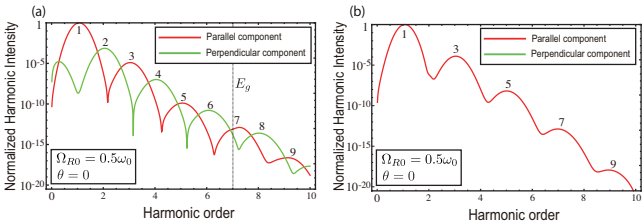}
\caption{High-harmonic spectra generated from (a) TMDCs and (b) graphene for the Rabi frequency $\Omega_{R0}=0.5 \omega_{0}$ and the angle $\theta=0$, where $\omega_0$ and $E_{g}$ are the frequency of the incident electric field and the bandgap energy. The red and green lines show the parallel and perpendicular emissions with respect to the incident electric fields, respectively.}
\end{figure}

For the derivation of Hamiltonian of the system, we will apply the same procedure used in Refs. \cite{Tamaya1} and \cite{Tamaya2} to this model. Only considering nearest-neighbor hopping of electrons and employing the Coulomb gauge \cite{Haug}, we can arrive at a tight-binding Hamiltonian $H=H_{0}+H_{I}$, where
\begin{eqnarray}
{H}_0 &=& \sum_{\bm{k}} \left[\gamma f(\bm{k}) a_{\bm{k}}^\dagger b_{\bm{k}}+\gamma f^*(\bm{k}) b_{\bm{k}}^\dagger a_{\bm{k}}+m \left( a_{\bm{k}}^\dagger a_{\bm{k}} -b_{\bm{k}}^\dagger b_{\bm{k}} \right)\right], \nonumber 
\\ \nonumber
\\
{H}_{I} &=& -\hbar \sum_{\bm{k}} \left[\Omega_R (\bm{k},t) a_{\bm{k}}^\dagger b_{\bm{k}}+\mbox{c.c.}\right]. \nonumber
\end{eqnarray}
Here, $\gamma$ is the transfer integral, $\hbar$ is the Planck constant, $f(\bm{k})$ is a form factor defined as $f(\bm{k})=\sum_{i} e^{i \bm{k} \cdot \bm{\delta_{i}}}=| f(\bm{k}) |e^{i \theta_{f(\bm{k})}}$, $\bm{\delta}_{i}$ is a lattice vector, $a_{\bm{k}}(b_{\bm{k}})$ is the annihilation operator of electrons with wavenumber $\bm{k}$ on the atom A (B), and $\Omega_{R}(\bm{k},t)$ is the Rabi frequency defined by $\Omega_{R}(\bm{k},t)=(e \hbar/m_{0}c)\sum_{i} e^{i \bm{k} \cdot \bm{\delta}_{i}} \int d^2x \phi^{\ast}_B (\bm{x}) \bm{A}(t) \cdot \bm{p} \phi_A(\bm{x}-\bm{R}_{i})$, where $m_{0}$ is the electron mass, $e$ is the electron charge,  $c$ is the velocity of light, ${\bm{A}}(t)$ is the vector potential of the incident electric fields, and ${\bm{p}}$ is the momentum of the bare electrons. In this model, we will take the lattice vectors to be $ {\bm{\delta}_{1}}=(0,0)$, $ {\bm{\delta}_{2}}=(a/2,\sqrt{3}a/2) $, and $ {\bm{\delta}_{3}}=(a/2, -\sqrt{3}a/2) $, respectively, which represent Dirac points as $K^\pm$=$(\pm 4\pi/3a,0)$. Here, $a$ is the lattice constant. The expression of the Rabi frequency, whose definition is generally described as the product of the dipole moment ${\bm{d}}$ and the vector potential ${\bm{A}}(t)$ \cite{Haug}, certainly involves in information on both the lattice vectors and atomic orbitals. Only focusing near $K^{\pm}$ points and supposing the vector potential to be $\bm{A}(t)=A_{0}(\epsilon^2_{1}+\epsilon^2_{2})^{-1/2} \exp (-(t-t_{0})^2/T^2) (\epsilon_{1} \cos\omega_{0}t ,\epsilon_{2} \cos\omega_{0}t )$, we can approximate the Rabi frequency as $\Omega_{R}(\bm{k},t) ={\rm{Re}}[\Omega_{R}(\bm{k},t)]+i {\rm{Im}}[\Omega_{R}(\bm{k},t)]\approx \Omega_{R0}(t)[[\epsilon_{1}(\tau^z + \tilde{\beta} k_{x} a)-\epsilon_{2} \tilde{\gamma} k_{y} a] \cos\omega_{0}t+i \tau^z [\epsilon_{1} \tilde{\gamma} k_{y} a+\epsilon_{2}(\tau^z+\tilde{\beta} k_{x} a)] \cos \omega_{0}t$], where $\tilde{\beta}=-0.87$ and $\tilde{\gamma}=0.27$ are dimensionless constants calculated from the atomic orbitals and $\Omega_{R0}(t)=\Omega_{R0} \exp[-(t-t_{0})^{2}/T^2]$. Here, we define the time-independent Rabi frequency as $\Omega_{R0}$. Throughout the paper, we set $t_{0}=12 \pi /\omega_{0} $ and $T=4 \pi /\omega_{0}$. In our numerical calculation, instead of rotating crystals, we change the orientation angle of the incident electric field $\theta$ from 0 to $2\pi$, where $\tan \theta=(\epsilon_{1}/\epsilon_{2})$.

\begin{figure*}[tp] 
\centering
\includegraphics[width=.99\textwidth, clip]{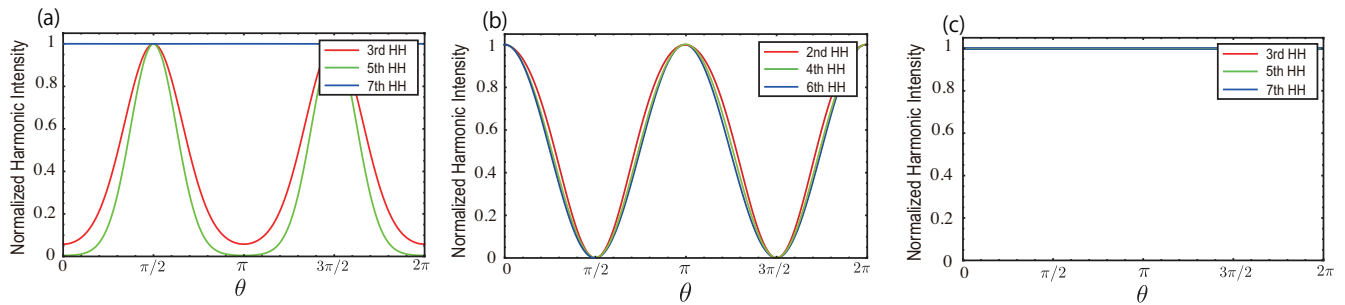} 
\caption{Orientation dependence of parallel (a) and perpendicular (b) emissions of HHG generated from TMDCs for $\Omega_{R0}=0.05\omega_{0}$. (c) Orientation angle of parallel emissions of HHG generated from graphene for $\Omega_{R0}=0.05\omega_{0}$. Red, green, and blue lines correspond to the 3rd, 5th, and 7th harmonics.}
\end{figure*}

The transformation of the Hamiltonian from the tight-binding to the band-structure picture can be performed by diagonalizing the single-particle part $H_{0} $ with the unitary transformation $a_{\bm{k}}=\alpha e_{\bm{k}}-\beta h^{\dagger}_{-\bm{k}} $ and $b_{\bm{k}}=-(\alpha' e_{\bm{k}}-\beta' h^{\dagger}_{-\bm{k}})$, where $e_{\bm{k}}(h_{\bm{k}})$ is the annihilation operator of electrons (holes) and the coefficients are given by $\alpha=[\gamma^{2} | f(\bm{k}) |^{2} /2E_{\bm{k}} (E_{\bm{k}}-m)]^{1/2}$, $\beta=[\gamma^{2} | f(\bm{k}) |^{2} /2E_{\bm{k}} (E_{\bm{k}}+m)]^{1/2}$, $\alpha'=-(E_{\bm{k}}-m)^{1/2}/(2E_{\bm{k}})^{1/2} e^{i \theta_{f({\bm{k}})}}$, and $\beta'=(E_{\bm{k}}-m)^{1/2}/(2E_{\bm{k}})^{1/2} e^{i \theta_{f({\bm{k})}}}$, respectively, where $E_{\bm{k}}=[\gamma^{2} | f(\bm{k}) |^{2}+m^{2} ]^{1/2}$. Only focusing near $K^\pm$ points, we can approximate the form factor $ f(\bm{k})$ as $ \gamma | f(\bm{k}) | \approx \hbar v_{F} (k_{x}-i \tau^{z} k_{y}) $ and $e^{i \theta_{ f( \bm{k} ) } } \approx e^{i \theta_{\bm{k}} }$, where $v_{F}$ is the Fermi velocity of graphene and $\theta_{\bm{k}} \equiv \arctan (k_y/k_x)$. Using this transformation, we can derive a Hamiltonian of the form $H=H_{0}+H_{I}$, where
\begin{eqnarray}
{H}_0 &=& \sum_{\bm{k}} E_{\bm{k}} (e^{\dagger}_{\bm{k}}e_{\bm{k}}+h^{\dagger}_{-\bm{k}}h_{-\bm{k}}) , \\
{H}_{I} &=& \hbar \sum_{\bm{k}} [(\hbar v_{F} k/E_{\bm{k}} )(\tau^{z} \cos \theta_{\bm{k}} {\rm{Re}}[\Omega_{R}(\bm{k},t)] \nonumber \\ 
&-&\sin \theta_{k} {\rm{Im}} [\Omega_{R} (\bm{k},t)])(e^{\dagger}_{\bm{k}}e_{\bm{k}}+h^{\dagger}_{-\bm{k}}h_{-\bm{k}}-1) \nonumber \\ [6pt]
&+&(m/E_{\bm{k}})(\tau^{z} \cos \theta_{\bm{k}} {\rm{Re}}[\Omega_{R} (\bm{k},t)] \nonumber \\ [6pt]
&-&\sin \theta_{\bm{k}} {\rm{Im}} [\Omega_{R}(\bm{k},t)]) (e^{\dagger}_{\bm{k}}h^{\dagger}_{-\bm{k}}+h_{-\bm{k}}e_{\bm{k}}) \nonumber \\ [6pt]
&+&i \tau^{z} (\tau^{z} \sin \theta_{k} {\rm{Re}}[\Omega_{R} (\bm{k},t)] \nonumber \\ [6pt]
&+&\cos \theta_{\bm{k}} {\rm{Im}} [\Omega_{R}(\bm{k},t)]) (e^{\dagger}_{\bm{k}}h^{\dagger}_{-\bm{k}}-h_{-\bm{k}}e_{\bm{k}}). 
\end{eqnarray}
The bandgap energy of the system can be estimated as $E_{g}=2m$ from the representation $E_{\bm{k}}=[ (\hbar v_{F} k )^2+m^{2} ]^{1/2}$. Utilizing this Hamiltonian, the time evolution equations of the densities $ f_{\bm{k}}^{\sigma}=\langle \sigma_{\bm{k}}^{\dagger} \sigma_{\bm{k}} \rangle$ and polarization $ P_{\bm{k}}= \langle h^{\dagger}_{-\bm{k}}e_{\bm{k}} \rangle$ with Bloch wavevector $\bm{k}$ can be derived as
\begin{eqnarray}
i \hbar \frac {\partial}{\partial t} P_{\bm{k}}&=& 2 \left[ \epsilon_{\bm{k}}^{e}(t)+\epsilon_{\bm{k}}^{h}(t) \right] P_{\bm{k}}+\hbar \left(m/E_{\bm{k}} \right) \nonumber \\ [6pt]
&&\times(\tau^{z} \cos \theta_{\bm{k}}{\rm{Re}}\left[\Omega_{R}(\bm{k},t)\right]-\sin \theta_{\bm{k}} {\rm{Im}} \left[ \Omega_{R} ( \bm{k},t) \right]) \nonumber \\ [6pt] 
&& \times \left[1-f^{e}_{\bm{k}}-f^{h}_{\bm{k}}\right] \nonumber \\ [6pt]
&+&i \hbar (\sin \theta_{\bm{k}} {\rm{Re}}[\Omega_{R}(\bm{k},t)]+\tau^{z} \cos \theta_{\bm{k}} {\rm{Im}} [\Omega_{R} (\bm{k},t)]) \nonumber \\ [6pt]
&&\times [1-f^{e}_{\bm{k}}-f^{h}_{\bm{k}}]-i \gamma_{t} P_{\bm{k}}, \\ [6pt] \nonumber
\frac {\partial}{\partial t} f_{\bm{k}}^{\sigma}&=&-2 \tau^{z} [(\tau^{z} \sin \theta_{\bm{k}} {\rm{Re}}[\Omega_{R}(\bm{k},t)] \nonumber \\ [6pt]
&& +\cos \theta_{\bm{k}} {\rm{Im}} [\Omega_{R} (\bm{k},t)]) {\rm{Im}}[(i P_{\bm{k}})^{\dagger}] \nonumber \\ [6pt]
&&+2(m/E_{\bm{k}})[(\tau^{z} \cos \theta_{\bm{k}} {\rm{Re}}[\Omega_{R}(\bm{k},t)] \nonumber \\ [6pt]
&& -\sin \theta_{\bm{k}} {\rm{Im}} [\Omega_{R} (\bm{k},t)]){\rm{Im}}[(P_{\bm{k}})^{\dagger}]-\gamma_{l}f_{\bm{k}}^{\sigma}.
\end{eqnarray}
Here, $\gamma_{t} $ and $\gamma_{l} $ are the transverse and longitudinal relaxation constants, and here, they are fixed to $ \gamma_{t}=0.1 \omega_{0}$ and $ \gamma_{l}=0.01 \omega_{0}$ \cite{Tamaya1}. The numerical solutions of these equations give the time evolutions of the distributions of the carrier densities $f_{\bm{k}}^{\sigma}$ and polarization $P_{\bm{k}}$ in two-dimensional $\bm{k}$ space. Utilizing these numerical solutions of $P_{\bm{k}}$ and $f_{\bm{k}}^{\sigma}$, the time evolutions of the generated current $J(t)=(J_{x}(t),J_{y}(t))$ along the $x$- and $y$-axes can be calculated on the basis of $J_{\nu}(t)=-c\langle \partial H_{I}/\partial A_{\nu} \rangle (\nu=x,y)$. The parallel and perpendicular components of $J(t)$ with respect to the incident electric field are given by $J_{\parallel}(t)=J_{x}(t) \cos \theta+J_{y}(t) \sin \theta$ and $J_{\perp}(t)=-J_{x}(t) \cos \theta+J_{y}(t) \sin \theta$, respectively. Then, the high-harmonic spectra can be calculated on the basis of $I_{\sigma}(\omega)=\omega^2 |J_{\sigma}(\omega)|^2$, where $J_{\sigma}(\omega)$ is the Fourier transform of the current vector $J_\sigma(t)$ with $\sigma=\parallel, \perp$. Below, we compare the characteristics of HHG spectra in TMDCs and graphene ($\tilde{\beta}=\tilde{\gamma}=0$ and $m$=0) \cite{Stroucken,Yoshikawa}, supposing the bandgap energy in TMDCs to be $E_{g}=7 \hbar \omega_{0}$ \cite{Liu,Yoshikawa}.

Figures 2(a) and (b) show the higher-order harmonic spectra generated from TMDCs and graphene, respectively, in the case of $\Omega_{R0}=0.5\omega_{0}$ and $\theta=0$. Here, the parallel and perpendicular components of HHG with respect to the incident electric field are plotted as red and green lines. These figures clearly show that both parallel and perpendicular emissions exist in TMDCs and they respectively involve odd- and even-order harmonics, while in graphene, only the parallel emission exists and it involves odd-order harmonics.

This difference in HHG between TMDCs and graphene can be qualitatively explained in terms of the symmetries of the dipole moments ${\bm{d}}$, which is involved in the Rabi frequency $\Omega_{R}(\bm{k},t)=(e \hbar/m_{0}c)\sum_{i} e^{i \bm{k} \cdot \bm{\delta}_{i}} \int d^2x \phi^{\ast}_B (\bm{x}) \bm{p} \cdot \bm{A}(t) \phi_A(\bm{x}-\bm{R}_{i}) \propto {\bm{d}} \cdot {\bm{E}}(t)$. TMDCs and graphene have the same lattice vector ${\bm{\delta_{i}}}$, but different symmetries of atomic orbitals. In the case of TMDCs, the orbital wavefunction is given by $\phi_A(\bm{x})=p_x(\bm{x})+ i p_y (\bm{x})$ and $\phi_B(\bm{x})=d_{xy} (\bm{x})+i \tau^z d_{x^2-y^2} (\bm{x})$, while in graphene, it is given by $\phi_A(\bm{x})=\phi_B(\bm{x})=\phi_{2p_z}(\bm{x})$. Near $K^{\pm}$ points, the dipole moments can be approximated as ${\bm{d}}=(d_{x},d_{y}) \approx (d_{x},i\tau^z d_{x}) \propto (\tau^z+\tilde{\beta}k_{x}+i \tilde{\gamma}\tau^z k_{y},i \tau^z d_{x})$ for TMDCs and ${\bm{d}} \propto (\tau^z,i) $ for graphene, respectively. By assuming ${\bm{E}}(t)=(E_{x}(t),0)$, the Rabi frequency is given by $\Omega_{R}(\bm{k},t) \propto d_{x}E_{x}= [\tau^z + \tilde{\beta} k_{x}+i \tau^z \tilde{\gamma} k_{y}] E_{x}(t) $ for TMDCs and $\Omega_{R}(\bm{k},t)\propto \tau^z E_{x}(t)$ for graphene. The real and imaginary parts of the Rabi frequency $\Omega_{R}(\bm{k},t)$ can be related to the parallel and perpendicular components of HHG~\cite{Tamaya2}. Thus, the Rabi frequency for TMDCs has both parallel and perpendicular components of HHG. Moreover, the parallel and perpendicular components of the dipole moment $d_{x}$ are, respectively, variant and invariant under a space inversion, $i. e.$, $k_{x} \to -k_{x}$, $k_{y} \to -k_{y}$, and $\tau^{z} \to -\tau^{z}$. Considering a relationship $P_{\parallel} \propto {\rm{Re}}[d_{x}]$ and $P_{\perp} \propto {\rm{Im}}[d_{x}]$ \cite{Loudon}, where $P_{\parallel}$ and $P_{\perp}$ are the polarizations in  parallel and perpendicular directions with respect to the incident electric field, we can show that the even (odd)-order susceptibility vanishes for parallel (perpendicular) components \cite{Boyd,Shen,Yariv}. Therefore, the parallel (perpendicular) emission of HHG for TMDCs has only odd (even)-order harmonics. On the other hand, in graphene, the Rabi frequency does not include the wavenumber ${\bm{k}}$ and only has the parallel component of HHG with only odd-order harmonics.


Figures 3(a) and (b) show the orientation dependence of the parallel and perpendicular emissions of HHG generated from TMDCs in the case of $\Omega_{R0}=0.05\omega_{0}$. Figure 3(a) indicates that the third (red line) and fifth (green line) harmonics of the parallel emission have $\pi$ periodicity as a function of the orientation angle $\theta$. On the other hand, the seventh (blue line) and ninth (not shown) harmonics have almost no dependence on the orientation of crystal, which is consistent with recent experiments \cite{Liu}. Figure 3(b) shows that the second, fourth, and sixth harmonics of the perpendicular emissions also have $\pi$ periodicity with a $\pi/2$ phase shift compared with the parallel emissions.


\begin{figure}[t] 
\includegraphics[width=8.5cm]{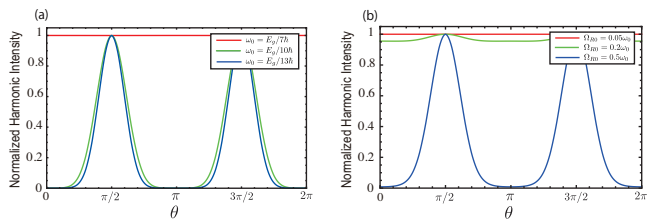}
\caption{(a) Orientation angle $\theta$ dependence of the parallel emission of HHG from TMDCs for $\Omega_{R0}=0.05\omega_{0}$. Red, green, and blue lines correspond to the $\theta$ dependence for different frequencies $\omega_0$ of the incident field. (b) The orientation angle $\theta$ of the parallel emission of HHG generated from TMDCs for $E_{g}=7 \hbar \omega_{0}$. Red, green, and blue lines correspond to different intensities of the incident electric field.
}
\end{figure}

To understand these behaviors, we plot in Fig.~3(c) the orientation dependence of the parallel emissions in graphene as a reference. In contrast to TMDCs, all the odd harmonics have no orientation dependence. The difference in periodicity between TMDCs and graphene can be qualitatively explained from the dipole moments ${\bm{d}} \approx (d_{x},i\tau^z d_{x}) \propto (\tau^z+\tilde{\beta}k_{x}+i \tilde{\gamma}\tau^z k_{y},i \tau^z d_{x})=(\tau^z+\tilde{\beta}k \cos \theta_{k}+i \tilde{\gamma}\tau^z k \sin \theta_{k},i \tau^z d_{x})$ for TMDCs and ${\bm{d}} \propto (\tau^z,i) $ for graphene. These expressions indicate that the $\pi$ periodicity for TMDCs comes from the $\cos \theta_{\bm{k}}$ and $\sin \theta_{\bm{k}}$ terms. In addition, the relative $\pi/2$ phase difference between parallel and perpendicular emissions for TMDCs can be explained by the relation $d_{y}\propto i d_{x} = e^{i \pi/2} d_x$. Note that the dipole moments in graphene (${\bm{d}} \propto (\tau^z,i) $) explicitly have no $\theta$ dependence, clearly indicating the parallel emissions from graphene are constant as a function of $\theta$. Thus, we conclude that the periodic $\theta$ dependence of HHG is related to that of the dipole moments ${\bm{d}}$, which is determined by the symmetries of the atomic orbitals and the crystal structure. Note as well that $\pi/3$ periodicity for HHG from TMDCs has been observed experimentally~\cite{Kumar, Malard, Zeng2,Li,Liu}. The difference in periodicity between the numerical and experimental results could be attributed to the atomic orbitals contributing to the HHG process. Therefore, we expect $\pi$ periodicity to be explicitly observed in the case of resonant HHG excitation only at $K^{\pm}$ points, as considered in this paper.

Finally, let us discuss the physical origin of the $\theta$ dependence of the seventh harmonic (Fig. 3(a))~\cite{Liu}. In the numerical calculation, we set the bandgap energy to $E_{g}=7\hbar \omega_{0}$, and as a result, almost no $\theta$ dependence of the seventh HHG process was attributed to the resonance excitation. To see the influence of the detuning on the seventh HHG from the resonant condition ($E_{g}=7\hbar \omega_{0}$), in Fig. 4(a), we plot the detuning dependence of the seventh harmonics for $\omega_{0}=E_{g}/7\hbar$ (red line), $\omega_{0}=E_{g}/10\hbar$ (green line), and $\omega_{0}=E_{g}/13\hbar$ (blue line) with $\Omega_{R0}=0.05\omega_{0}$. The result shows that the $\pi$ periodic $\theta$ dependence of the seventh harmonic becomes clear with increasing detuning. Therefore, we can observe the $\pi$ periodic $\theta$ dependence of the HHG spectrum in strong fields, because high-intensity electric fields give rise to temporal variations in the energy band structure and the concept of a static energy bandgap is irrelevant~\cite{Tamaya1, Tamaya2}. Finally, we plot in Fig. 4(b) the intensity dependence of the seventh harmonic in TMDCs for $ \Omega_{R0}=0.05\omega_{0} $ (red line), $ \Omega_{R0}=0.2\omega_{0} $ (green line), and $ \Omega_{R0}=0.5\omega_{0} $ (blue line) with the resonant condition $E_{g}=7\hbar \omega_{0}$. As expected, the $\pi$ periodic $\theta$ dependence becomes pronounced with increasing field intensity, indicating that the dynamics of the energy bandgap are essential to understanding the $\pi$-periodic $\theta$-dependence of HHG.


In conclusion, we theoretically investigate the orientation dependence of HHG in monolayer TMDCs. We found that TMDCs show both perpendicular and parallel emissions with respect to the incident electric field and they principally have only odd- (for perpendicular) or even-order (for parallel) harmonics. Moreover, the orientation dependence of the perpendicular and parallel emissions show a $\pi$-periodicity with a $\pi/2$ relative phase difference. These characteristics are quite different from those of graphene and conventional solid-state materials, and their anomalous behaviors can be attributed to the ${\bm{k}}$-dependence of the dipole moments, including the symmetries of the atomic orbitals and crystal structures in solid-state materials. This consideration indicates that the orientation dependence of HHG is mainly dominated by the interband excitation processes of Bloch electrons. Our results are qualitatively consistent with recent experimental results \cite{Kumar, Malard, Zeng2,Li,Liu} and provide a possible way to develop high-harmonic spectroscopy for solid-state materials.


\section*{Acknowledgements}
This work was supported by JST CREST (JPMJCR14F1), JST Nanotech CUPAL, and MEXT KAKENHI(15H03525).


\end{document}